\documentclass{sig-alternate-10pt}

\usepackage{graphicx}
\usepackage{caption}
\usepackage{subcaption}
\usepackage{color}
\usepackage{epstopdf}
\usepackage{url}
\usepackage{cite}
\begin{document}
%\title{Exploring use of network coding in multi-interface radio network}
\title{HetNetwork Coding: Scaling Throughput in Heterogeneous Networks using  Multiple Radio Interfaces}

\numberofauthors{1}
\author{
\alignauthor
Ratnesh Kumbhkar, Narayan Mandayam, Ivan Seskar\\ 
       \affaddr{WINLAB, Rutgers University,}
       \affaddr{671 Route 1 South,}\\
       \affaddr{North Brunswick, NJ, USA}\\
       \email{\{ratnesh, narayan, seskar\}@winlab.rutgers.edu}
       }
       
%\numberofauthors{1}
%\author{
%\alignauthor
%Omitted for submission\\
%Paper \# 255
%       }

\maketitle

\begin{abstract}
 The explosive demand for data has called for solution approaches that range from spectrally agile cognitive radios with novel spectrum sharing, to use of higher frequency spectrum as well as smaller and denser cell deployments with diverse access technologies, referred to as heterogeneous networks (HetNets). Simultaneously, advances in electronics and storage, has led to the advent of wireless devices equipped with multiple radio interfaces (e.g. WiFi, WiMAX, LTE, etc.) and the ability to store and efficiently process large amounts of data. Motivated by the convergence of HetNets and  multi-platform radios, we propose {\em HetNetwork Coding} as a means to utilize the available radio interfaces in parallel along with network coding to increase wireless data throughput. Specifically we explore the use of random linear network coding at the network layer where packets can travel through multiple interfaces and be received via multihoming. Using both simulations and experimentation with real hardware on WiFi and WiMAX platforms, we study the scaling of throughput enabled by such HetNetwork coding. We find from our simulations and experiments that the use of this method increases the throughput, with greater gains achieved for cases when the system is heavily loaded or the channel quality is poor. Our results also reveal that the throughput gains achieved scale linearly with the number of radio interfaces at the nodes.
\end{abstract}



 % NOT required for Proceedings

\section{Introduction}
Global data traffic has been increasing at an astounding rate and this trend is anticipated well into the future. There are many reasons for this,  such as increasing number of people with access to various types of  communication devices (e.g. laptop, tablet, smartphone etc), increase in traffic demand per user due to changing consumer behavior (e.g. the common consumption of and reliance on multimedia content), and increasing machine-to-machine data traffic. The wireless research community and industry at large are actively seeking out solutions that are needed to provide the capacity required to support the exploding volume of future wireless applications and services. In fact, there is a recognition and push in both industry and academia towards the goal of achieving ”1000x” capacity for wireless \cite{1000x, lte12bnsn, 1000xmeet}. The solution approaches range from spectrally agile cognitive radios with authorized shared access (ASA) spectrum sharing \cite{specsharing, 1000xmore}, to use of higher frequency spectrum \cite{current5G, rapp5g13} as well as smaller and denser cell deployments \cite{ping10femtoUL}. The result has been research and proposals on heterogeneous networks (HetNets) \cite{andrews13hetnets, andrews14load, zhang04madm}, and self-organizing networks (SONS) \cite{son4g, sohrabi99sensornet}. While there is no ``magic bullet'' to achieve the 1000x capacity in the HetNets, the technical approaches include advanced antenna design, use of higher frequency spectrum as well as better channel access and coordination methods to mitigate interference.  

\begin{figure}[t]
\begin{center}
\includegraphics[width=0.45\textwidth]{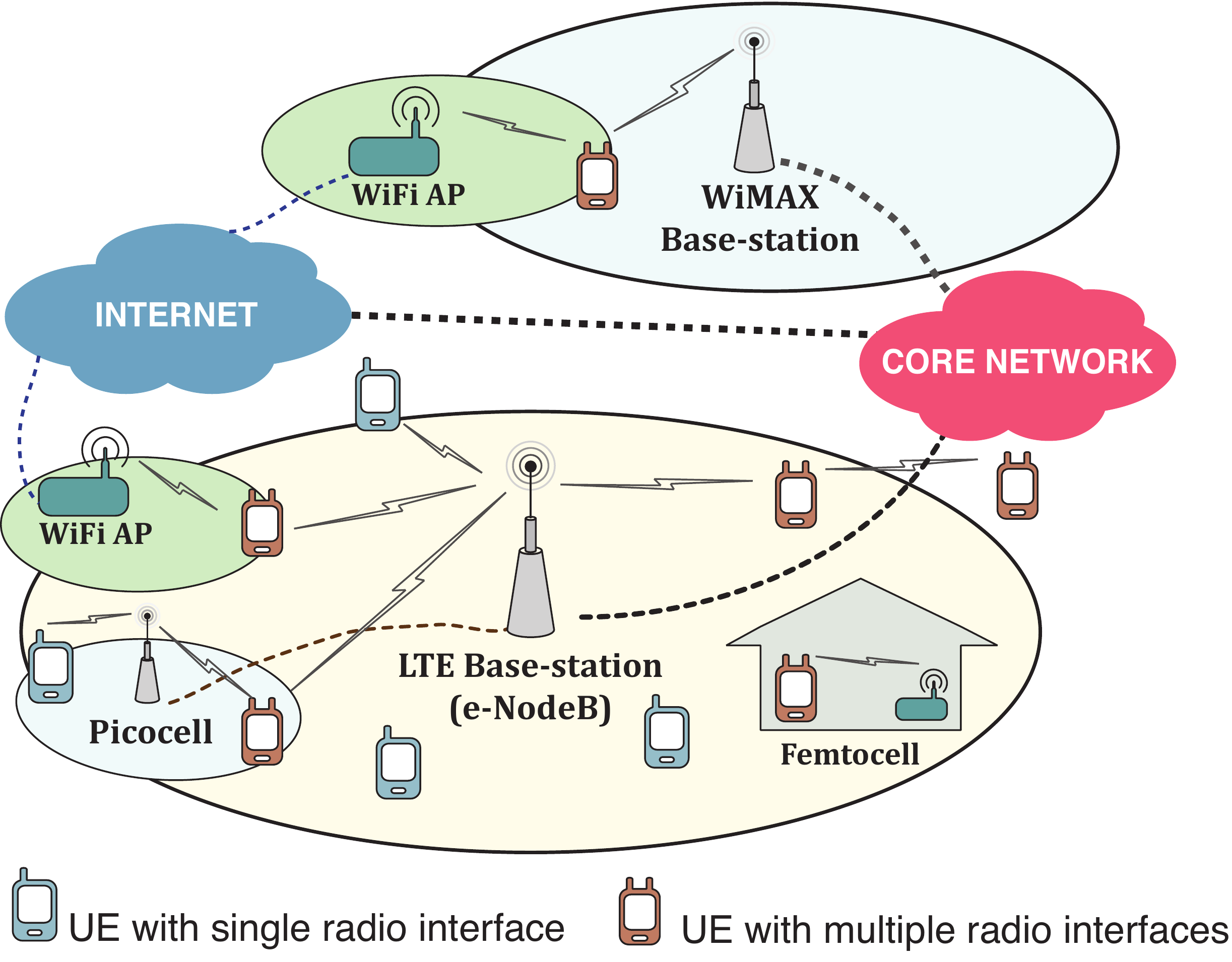}
\end{center}
\caption{An example of Heterogenous Network}
\label{fig:hetnet}
\end{figure}

In this paper, we bring a new dimension to scaling capacity in HetNets, by taking advantage of the multiplicity of radio platforms on a single device. Figure \ref{fig:hetnet} shows an example of a HetNet, which includes various WiFi hotspots, a WiMAX base-station and an LTE base-station, for which the coverage is increased by deploying some small cells.  As shown in the figure \ref{fig:hetnet}, many wireless devices are equipped with multiple radio interfaces, which are capable of communicating via different frequency bands and different access techniques. e.g. a normal smart phone can use WiFi, LTE and bluetooth technologies for data communication. Generally these interfaces are used independently, and even though they correspond to different MAC and  PHY layer,  they can be shared at the network layer. Therefore these interfaces provide an  opportunity for increasing the throughput  in HetNets. While multihoming has been around as a concept in IP networks \cite{Bu04multihoming,Sousa11multihoming}, it has never been explicitly considered in HetNets in the context of scaling throughput. We propose to use these different interfaces to form parallel networks, by either forming ad-hoc network or infrastructure based network to offload some portion of traffic. 
%An Example of HetNets is given by figure \ref{fig:hetnet}, which is centered around LTE architecture. base-station is denoted is e-NodeB (eNB), mobile node is UE, smaller cells are picocells or femtocells with their own base-stations. These base-stations are interconected by wired or wireless backhaul.   HetNets have already been shown to be a way to improve cellular system capacity and coverage, and it is one of the main potential approaches to achieve 1000x capacity goal \cite{Hu11hetnets}\cite{andrews13hetnets}.

The use of multiple interfaces at any node for a single communication session in a heterogenous network can result in complicated routing schemes, therefore we explore use of \emph{HetNetwork coding}, i.e. network coding in the heterogeneous network,  to mitigate this problem. Network coding is a technique, in which outgoing information (packet) at a node is coded as a function of incoming informations (packets) \cite{Ahlswede00networkinformation}. Network coding has shown advantages over traditional methods in terms of throughput, scalability and security in both wired and wireless networks. It has been shown that linear network coding is able to achieve capacity in a multicast network and choosing random coefficients from a sufficiently large finite field is sufficient for practical purposes \cite{Ho06arandom,Ho03onrandomized}. Since the use of random network coding allows packets to come from any path and in any order, as long as these packets are decodable, it becomes simple to handle routing of packets.

Motivated by the convergence of HetNets and multi-platform radios, in this paper we propose the use of {\em HetNetwork Coding}. Specifically we investigate benefits in a  scenario where wireless nodes have two interfaces i.e. cellular and WiFi. We evaluate the performance of this scheme using MATLAB simulations, when WiFi interfaces form an ad-hoc network or connect via an access point. We perform these simulations for varying cellular link quality  with cellular system loading. We also perform experiments  on the the ORBIT testbed \cite{Raychaudhuri05orbit} using a WiMAX base station and WiFi nodes for similar scenarios, showing the advantage of using network coding in the heterogenous network.

\section{Related work}
Heterogeneous networks have been studied for increasing the LTE throughput and coverage area by using a variety of cell sizes, access techniques and transmit powers \cite{Damn11surveyhetnet}\cite{ping11hetnet}. There are many technical challenges associated with HetNets such as resource allocation, interference, backhauling and handover among others and some of which is addressed by authors in \cite{lopez11hetnetinterference,haoming11fiberhetnet}. HetNets are being considered as the major solution to handle the huge data traffic demand in cellular networks and  methods to efficiently and effectively model these are discussed in  \cite{andrews13hetnets}.

A specific case of HetNets, where only two types of networks (mainly WiFi and cellular) are available, has been extensively studied. It has been shown by Gupta and Kumar \cite{Gupta00thecapacity} that for a wireless ad-hoc network with $n$ identical nodes, where interference has been modeled as protocol model, throughput (in bits per second) scales no better than $\frac{\sqrt{n}}{\sqrt{\log{n}}}$. Communication scenarios where a wireless ad-hoc network is used with a wired infrastructure (cellular backbone network) has been studied by Liu et. al in  \cite{towsley03capacity}, where they calculate the throughput of such a hybrid network. They show that, if the number of infrastructure nodes increase as $\Omega(n)$, then the throughput increase is better than the Kumar-Gupta bound. Similar cases  have been studied in detail in  \cite{Kulkarni03throughputscaling} and \cite{Reznik04asmall} in the context of scaling laws for   throughput of  wireless ad-hoc networks. In these papers authors study the scaling of ad-hoc wireless networks after adding long wired links between few nodes.  Authors in \cite{Kulkarni03throughputscaling} find that after adding a highly structured wired infrastructure with access points  on top of a wireless ad-hoc network, a  throughput of $O(\frac{\sqrt{n}}{\sqrt{\log{n}}}+\frac{b_n}{\log{b_n}})$ can be achieved, where $b_n$ is the number of base stations in wired infrastructure. It is shown in \cite{Kulkarni03throughputscaling} that the overall throughput can be increased from the Kumar-Gupta bound  for $b_n=\Omega(\sqrt{n})$ with distance of $O({n^{1/4}})$ between the base stations. Motivated by small world networks, Reznik et al. \cite{Reznik04asmall} further extend this study by placing the wired infrastructure randomly instead of placing it in a grid structure as used in \cite{Kulkarni03throughputscaling}.  Authors in \cite{Reznik04asmall} find that random placement of wired links based on an imposed probability law can further increase the throughput for wireless ad-hoc networks. Motivated by these theoretical results, we expect the idea of HetNetwork coding to be scalable and perform both simulation and experimental studies for performance evaluation.

There have been some previous studies related to offloading of cellular traffic on WiFi links, such as  \cite{dhh11}, \cite{lmrs07} where authors implement the offloading only at the destination cell by choosing a set of nodes, which can receive the data from the destination base station on behalf of the destination node and then forward it to the destination node via WiFi links. Wu et al. propose a system called iCAR (integrated cellular and ad-hoc relaying system) \cite{Wu01integratedcellular}, where ad-hoc relay nodes are strategically placed inside a cellular network to offload the traffic from congested cells to non-congested cells.  Hsieh et al. \cite{Hsieh:2002:UAN:513800.513805} suggest various schemes for using an ad-hoc network in a cellular packet data network, but in these schemes the base station is actively involved with the ad-hoc network in improving the performance. The scheme proposed in this paper does not require any explicit coordination between the cellular and WiFi networks, but instead relies on the multiplatform radio enabled wireless nodes to adapt their packet processing at the network layer.

We note that a very similar approach  to use multiple interfaces together with network coding has been proposed by Cloud et. al \cite{cloud13multitcp}. In this paper, the authors suggest the use of   network coding with a multi-path protocol based on Multi-path TCP (MPTCP), which is  currently at working group level of the IETF \cite{ietfmptcp}.  They  collect empirical  results for a heterogenous network in a mobile environment to reflect the performance gain due to the use of multiple interfaces and they create a model to give a mean-field approximation of throughput of MPTCP and MPTCP-NC (Multi-Path TCP with Network Coding). They use their empirical data with a theoretical model to compare MPTCP and MPTCP-NC, and show that MPTCP-NC performs better. Our work differs from Cloud's work in that, we not only give simulation results, but also provide experimental results showing  the benefits of  using network coding on multiple interfaces using real a WiMAX base-station and WiFi interfaces on the ORBIT testbed.
%calculate theoretical throughput for MPTCP/NC (Multi-Path TCP with Network Coding); we provide simulation and empirical result in heterogenous network while using network coding. We provide simulation results for  different cellular link conditions and present a comparative result to show conditions where network coding is advantageous. We also show throughput performance for such network by using network coding on actual radio testbed on some specific topologies. 

\section{System Model}
\begin{figure}[t]
\begin{center}
\includegraphics[width=0.45\textwidth]{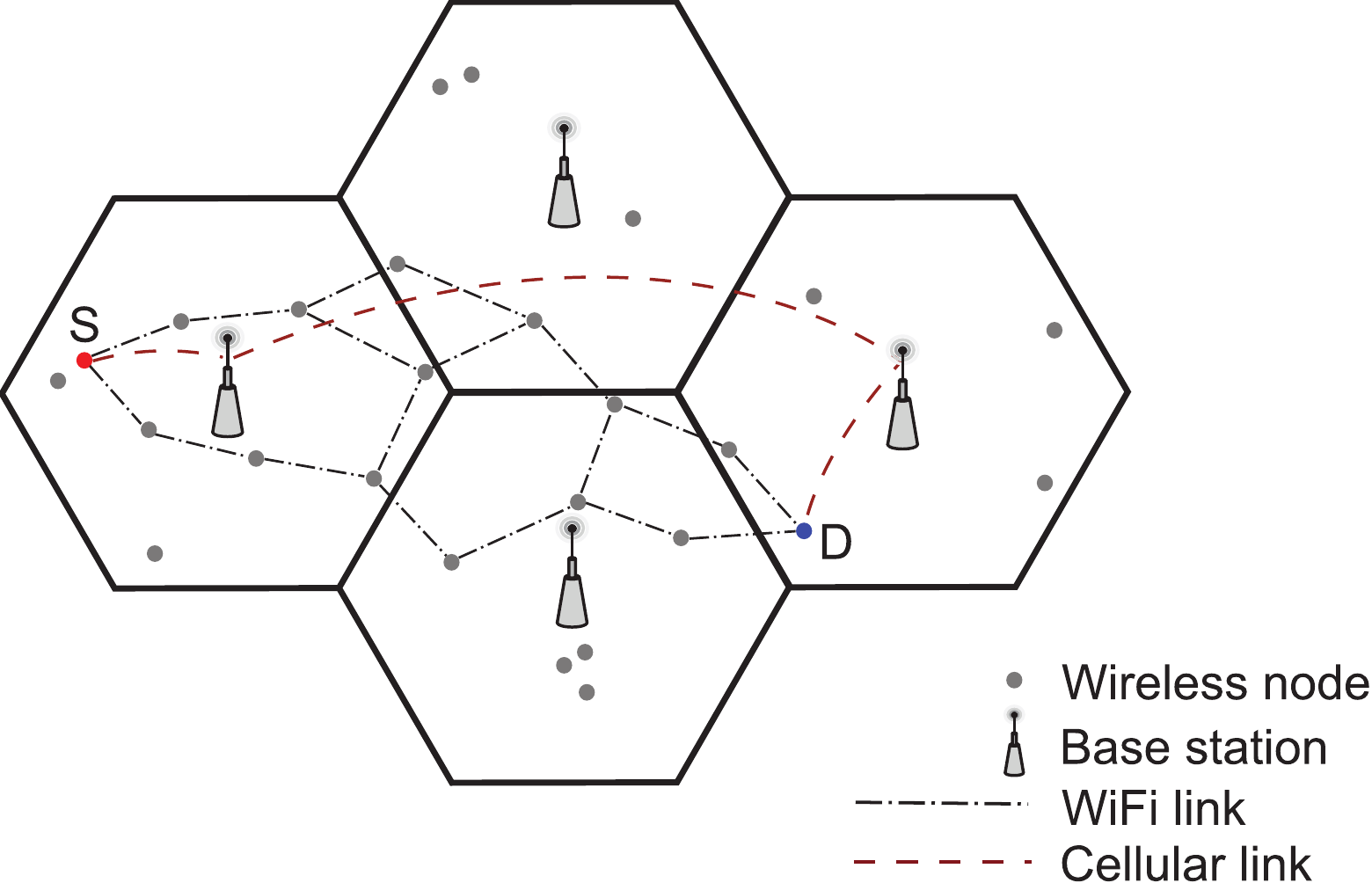}
\end{center}
\caption{A source destination pair (S-D)  in cellular network.}
\label{fig:7cell}
\end{figure}

In this paper, we illustrate the concept of HetNetwork Coding using the case of cellular and WiFi networks. We consider a cellular network with hexagonal cells  as shown in figure \ref{fig:7cell}.  Each cell  has a radius  $R$,  contains a base station at the center of the cell and can support a maximum data  rate of  $R_{cell}$ in both the uplink and downlink. We assume $n$ randomly placed nodes  within cellular network with uniform distribution, where each node is capable of simultaneously using both WiFi and cellular link for communication. For simplicity, we assume the following regarding the nodes in this HetNet:

\begin{itemize}
  \item All wireless nodes are similar, i.e. they use the same frequency, they transmit at the same power and have the same transmission rate, therefore all nodes have identical transmission range $r$.
  \item Nodes are capable of forming a wireless ad-hoc network using WiFi links.
  \item Nodes are either immobile or movement is relatively small  during the transmission period that channel conditions do not change significantly.
  \item Interference in WiFi link between  nodes is modeled by the so-called \emph{protocol model}.%, also known as the unit disk graph model.
\end{itemize}
%%%%
%%%%
 We assign a supported cellular  data rate  to each node based on the node's distance from its base station. Each wireless node contains a buffer of size $N_{buf}$ to store it's received packets. According to the protocol model if node $i$ is the transmitter and node $j$ is the intended receiver then this transmission will be successful only when
\begin{equation*}
d\{j,k\}\geq(1+\Delta)d\{i,j\} \;\; \forall k\neq i \; or \; j% \qedhere
\end{equation*}
where $d\{i,j\}$ is the euclidean distance between node $i$ and node $j$, and $\Delta$ is a positive number indicating the guard distance. The exact  choice of $\Delta$ depends on the  SNR  threshold necessary for successful packet reception. 
 We assume use of a single channel for WiFi communication. If two nodes are within the transmission range of each other, then they support rate $R_{WiFi}$. Communication between an S-D pair  shown in figure \ref{fig:7cell} is referred to as a  session. We assume that  there exists a  transport layer protocol for cellular communication which decides the data rate at the start of the session and maintains this rate throughout the entire session. The quality of the S-D link is decided by the worse of the 2 cellular links between (a) the source node and its corresponding base station or (b) the destination node and its corresponding base station. If the supported cellular data rates are $R_{cell,S}$ and  $R_{cell,D}$ at the source and destination node respectively, then the resulting S-D link throughput is given as:
 \begin{equation*}
R_{cell,link}=\min(R_{cell,S},R_{cell,D})
\end{equation*}

\newpage
\subsection{Network coding }
We assume that intra-session random network coding is used at the network layer, where the coding coefficients are elements chosen randomly from a finite field GF($2^h$), i.e. each coefficient contains $h$ bits.  Network coding is performed on a block of packets from the same session, where the block size is $M$. To decode these coded packets, the destination node can use one of the two approaches as follows. It can  check if each received packet is an innovative packet (i.e. the packet is linearly independent from previously received packets, therefore it can be used for decoding)  and only store innovative packets for decoding. Alternately, the destination node can receive a complete block of packets and then try to decode them. It is shown by Chou et. al \cite{Chou03practicalnetwork} that if random coefficients are chosen from a sufficiently large finite field, the block of packets can be decoded successfully with high probability. We choose $h=8$, which ensures that, at the receiver a full rank matrix can be formed after receiving $M$ packets with high probability.  If we represent an uncoded packet as a vector $\textbf{a}_i$ of length $k$ and the coefficients are represented as a matrix of size $M\times M$ with elements $c_{i,j}$, then encoding of the packet can be represented as 
 \begin{equation*}
 \begin{bmatrix}
   \textbf{b}_1 \\
   \textbf{b}_2 \\
   \vdots  \\
   \textbf{b}_M
 \end{bmatrix} =
 \begin{bmatrix}
  c_{1,1} & c_{1,2} & \cdots & c_{1,M} \\
  c_{2,1} & c_{2,2} & \cdots & c_{2,M} \\
  \vdots  & \vdots  & \ddots & \vdots  \\
  c_{M,1} & c_{M,2} & \cdots & c_{M,M}
 \end{bmatrix}
 \begin{bmatrix}
   \textbf{a}_1 \\
   \textbf{a}_2 \\
   \vdots  \\
   \textbf{a}_M
 \end{bmatrix}
 \end{equation*}
Note that the output vector $\textbf{b}_i$ is not the actual outgoing packet. Before sending the packet, coding coefficients $\{c_{i,1}, .... , c_{i,M}\}$ are appended to $\textbf{b}_i$. In our simulations and experiments we choose $M = 20$ bytes. Since, typically the size of an IP packet is $\approx 1400$ bytes, the overhead due to network coding is only  $1.428 \%$.

Figure \ref{fig:transfer} shows the end-to-end flow of packets for a source-destination pair. A source sends packets with a block ID on both WiFi and cellular links. Packets can arrive in different order based on the channel conditions of WiFi/cellular links or the number of hops on WiFi link. However it must be noted that the content of the received packets  can be very different from ones generated by the source because of possible network coding at intermediate nodes. As shown in figure \ref{fig:transfer}, we assume, once one block (i.e. $M$ packets) has been decoded by the destination node, it sends an acknowledgment to the source node for that block ID through an error free channel. Only when an acknowledgement has been received, the source node starts sending packets from the next block and this process continues.

\subsection{Routing}
We assume that packets are encoded at the source node and forwarded on both cellular and WiFi interfaces in parallel. We emphasize that at the source node the same copy of the packet is not sent on both the  WiFi and the cellular link, since this scheme would not utilize the ``multiplatform diversity'' available opportunistically, however it can provide more reliability. We assume that further network coding can be performed at the intermediate nodes for the same session but not at the base stations. 
We also assume that there exists a  routing protocol for routing the packets in the ad-hoc WiFi network such as, for example the one described in  \cite{cjkk06}. Once this routing protocol has converged, all nodes are aware of their respective distance in terms of the number of hops from other nodes. 
%Therefore any node which has higher hop distance to destination compared to previous node   automatically discards the packet to avoid flooding the network. 
If a node is on the right route then the received packet is stored in a buffer and if the node contains  other packets from the same session in the buffer, it uses the opportunity to perform intra-session network coding. This node creates a randomly coded packet from this session and sends it to its output ports. Any intermediate node has the following three choices while further forwarding the encoded packet  in the network:
\begin{figure}[t]
\begin{center}
\includegraphics[width=0.45\textwidth]{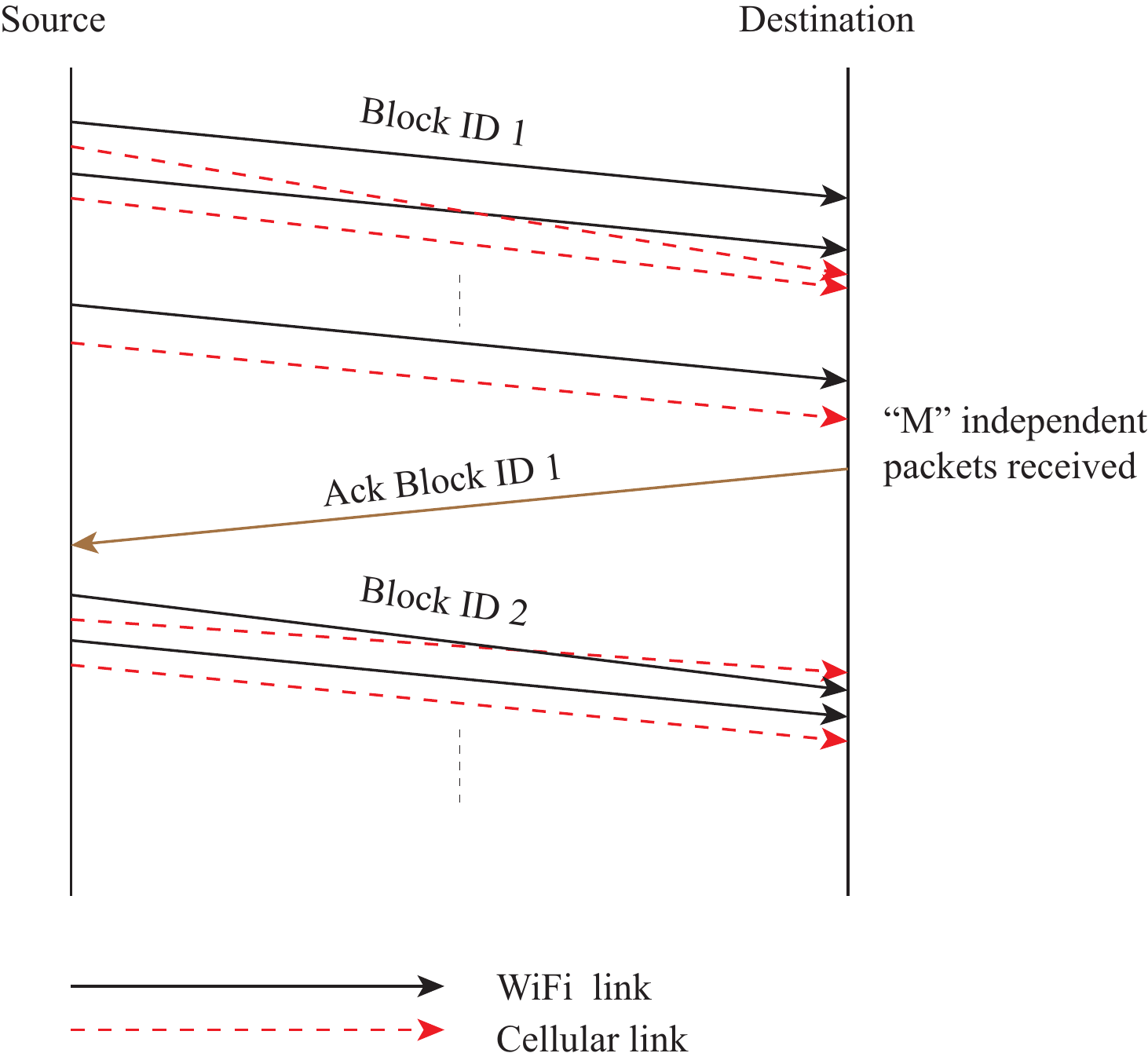}
\end{center}
\caption{End-to-end communication in a S-D pair}
\label{fig:transfer}
\end{figure} 

\begin{itemize}
  \item Send only using the cellular link : For cases where the destination has a better cellular link quality compared to the source node, an intermediate node with a good cellular link may choose to forward the packet via its cellular link. However use of the cellular link by the intermediate node will introduce addition load (traffic) in the cellular system.  
  \item Send only on WiFi : For cases where the destination's cellular link can support less data rate than the source's cellular link, it is not useful for the intermediate node to send the packet on  the cellular link hence it may choose to send data solely on WiFi.
  \item Send using both cellular and WiFi: The intermediate node can  send different  packets to its output interfaces in round-robin or in parallel or it can send the same copy to both interfaces or it can use any probabilistic scheme. 
\end{itemize}

We examine  all of these scenarios and identify the  possible solution in each of the scenarios.
%\subsection{Effect of loading}
We also study  the effect of loading on the cellular network due to the presence of multiple users in a cell. We consider the effect of loading using either equal rate for all the cell users or by allotting equal time to all the cell users.

\section{Performance evaluation}
To evaluate our proposed scheme,  we first  simulated it using MATLAB. To  obtain results under more realistic situations we also implemented it on the ORBIT testbed \cite{Raychaudhuri05orbit}. We use the ORBIT sandbox which consists of a WiMAX subsystem with base-station and controller, and nodes equipped with two wireless interfaces, namely WiMAX and WiFi. 

\subsection{Simulation}
\begin{figure}[t]
\begin{center}
\includegraphics[width=0.4\textwidth]{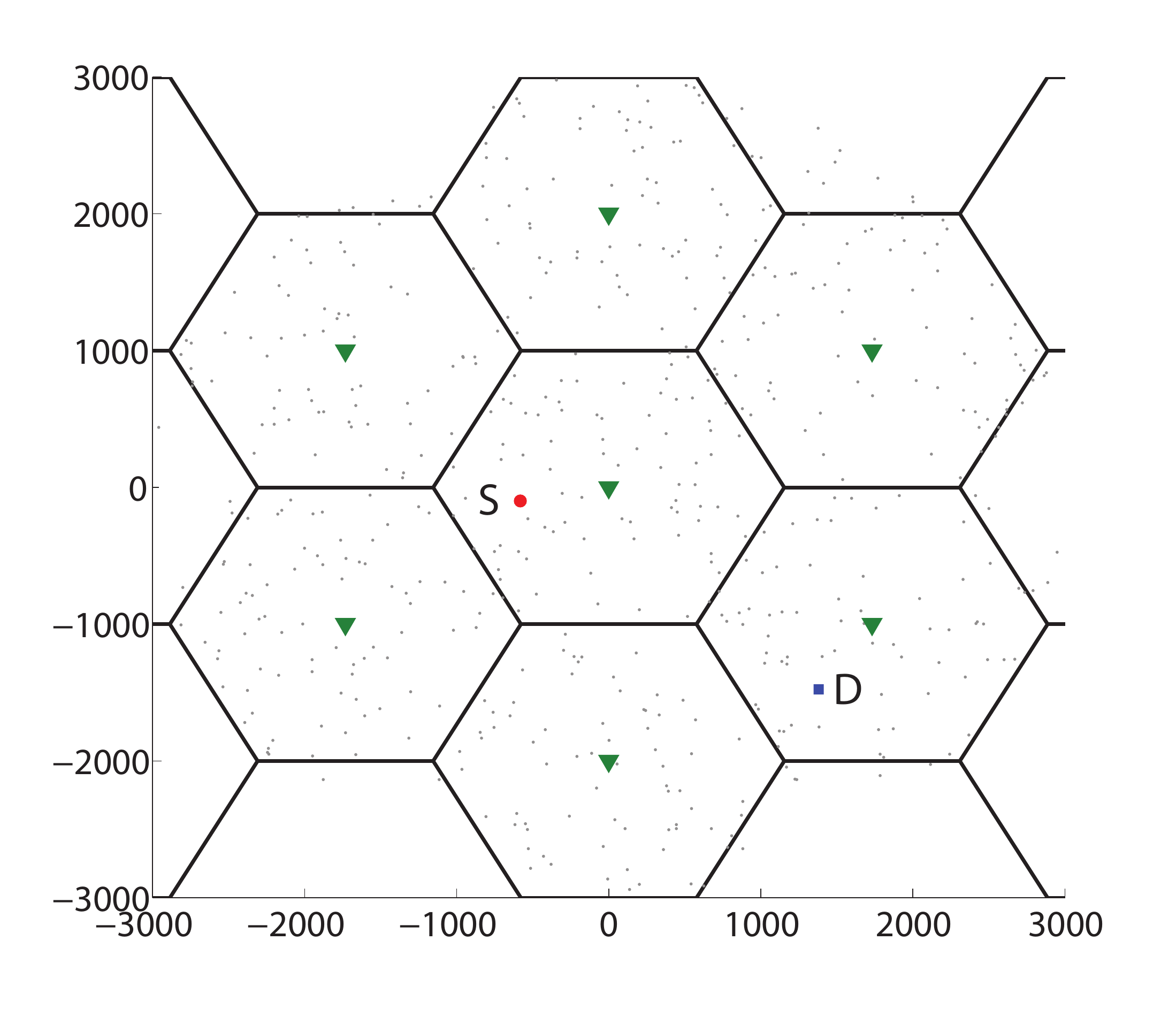}
\end{center}
\caption{Simulation setup}
\label{fig:simcell}
\end{figure}

Simulation on MATLAB is done using the communication system toolbox. We take a seven cell structure as shown in figure \ref{fig:simcell} for our simulation studies. Each hexagonal cell has a radius of $1000$ meters and we place approximately 700 to 800 nodes randomly inside this seven cell structure. Each node has a WiFi transmission range of 100 meters and the parameter $\Delta$ used for the protocol model  is  $0.2$.

We first consider the case of a  single session by choosing a random source destination pair in this network, with a constraint on the minimum number of hops. This constraint is placed to avoid the trivial case of a one-hop WiFi link. All  other nodes are available as  potential intermediate nodes for this communication session. At the intermediate nodes there is a choice of using  the cellular network with WiFi or just using WiFi. We simulate both situations and find that since at the destination, the link between the  base station and the destination node creates a bottleneck, use of the cellular network by intermediate nodes overloads the cellular network.  
We find that if we have a protocol where intermediate nodes forward the packets to the destination node using the cellular network, it affects the system in two ways (a) it increases the effect of loading in it's own cell (b) it creates congestion at the destination cellular link and the rate at which useful packets reach the destination node becomes really low. Due to these reasons, in our scheme intermediate nodes do not participate in cellular transfer during a communication session.   We compare our simulation results with the scenario where nodes can only communicate using the cellular link. We use  \emph{relative throughput} as the metric for performance, which is defined as the end-to-end throughput normalized by the single WiFi link throughput. We use this metric to compare the results from simulations with results obtained from experiments on the ORBIT testbed. We evaluate the simulation studies  for two different cases, first when the intermediate nodes can only form an ad-hoc network using their WiFi links, and second, when the intermediate nodes can form an infrastructure based WiFi network with access to a high capacity wired backbone network.
%\begin{figure}[h]
%\begin{center}
%\includegraphics[width=0.45\textwidth]{loadt.pdf}
%\end{center}
%\caption{Effect of loading with equal time sharing}
%\label{fig:loadt}
%\end{figure}
\subsubsection{Case 1: Ad-hoc WiFi network}
\begin{figure}[t]
%\begin{center}
\includegraphics[width=0.45\textwidth]{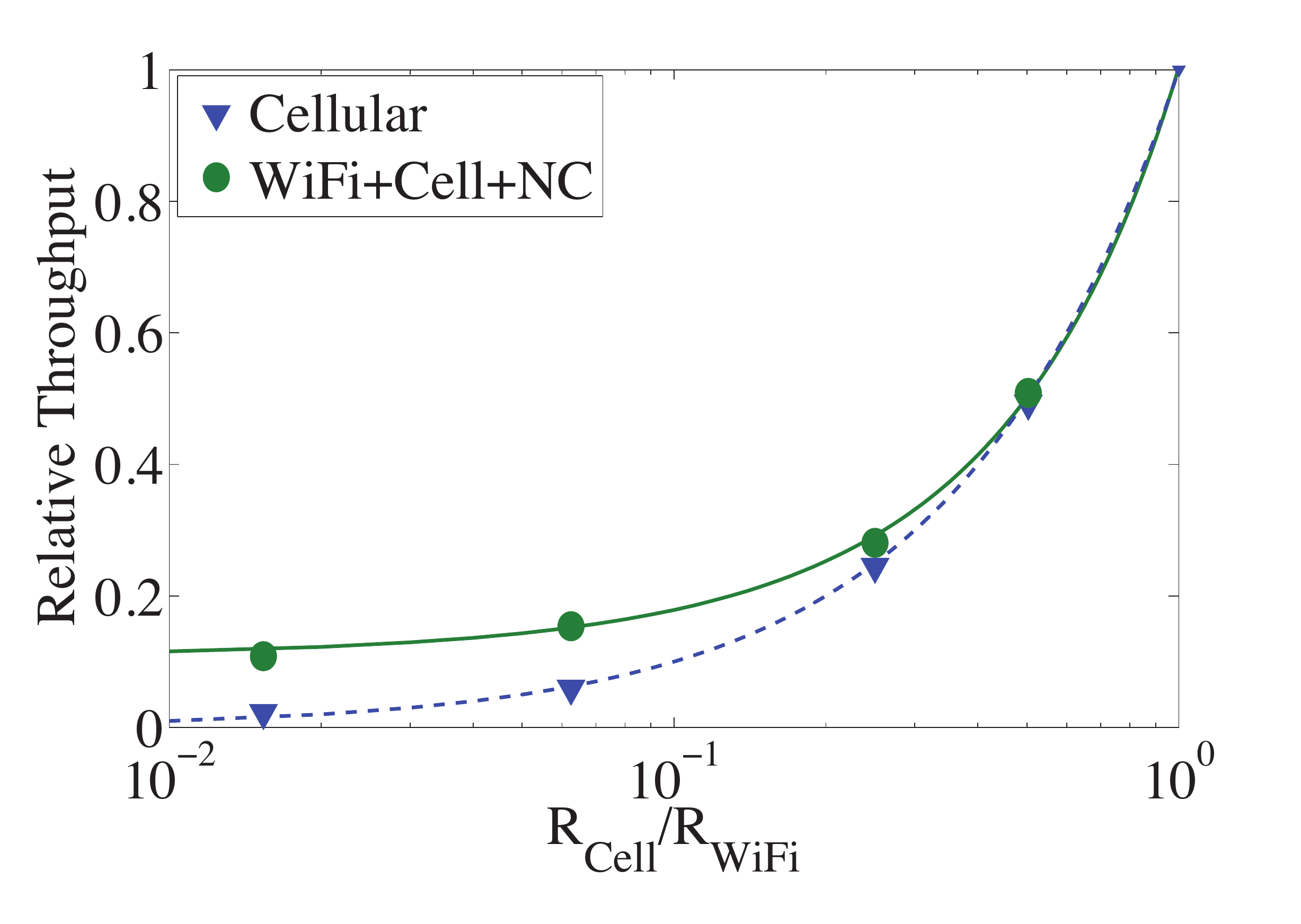}
%\end{center}
\caption{Relative throughput in multi-hop scenario}
\label{fig:rate}
\end{figure}

We find that sending network coded packets over the combination of WiFi and cellular  increases the average cellular throughput. Figure \ref{fig:rate} compares the average relative throughput as a function of varying cellular link quality between the cases where only the cellular link is used and where a combination of WiFi and cellular with network coding is used. We can see that when the cellular data rate is comparable  with the WiFi data rate, there is very little or no benefit of using network coding on both interfaces. This is because, under the protocol model for interference, the throughput of the multi-hop WiFi network decreases and saturates beyond some number of hops. Therefore almost all innovative packets for a block reach the  destination node through the cellular link directly from  the source. But as the cellular data rate becomes small compared to WiFi (e.g. towards the edge of the cell), using both interfaces provides greater gain.

\begin{figure}[t]
%\begin{center}
\includegraphics[width=0.45\textwidth]{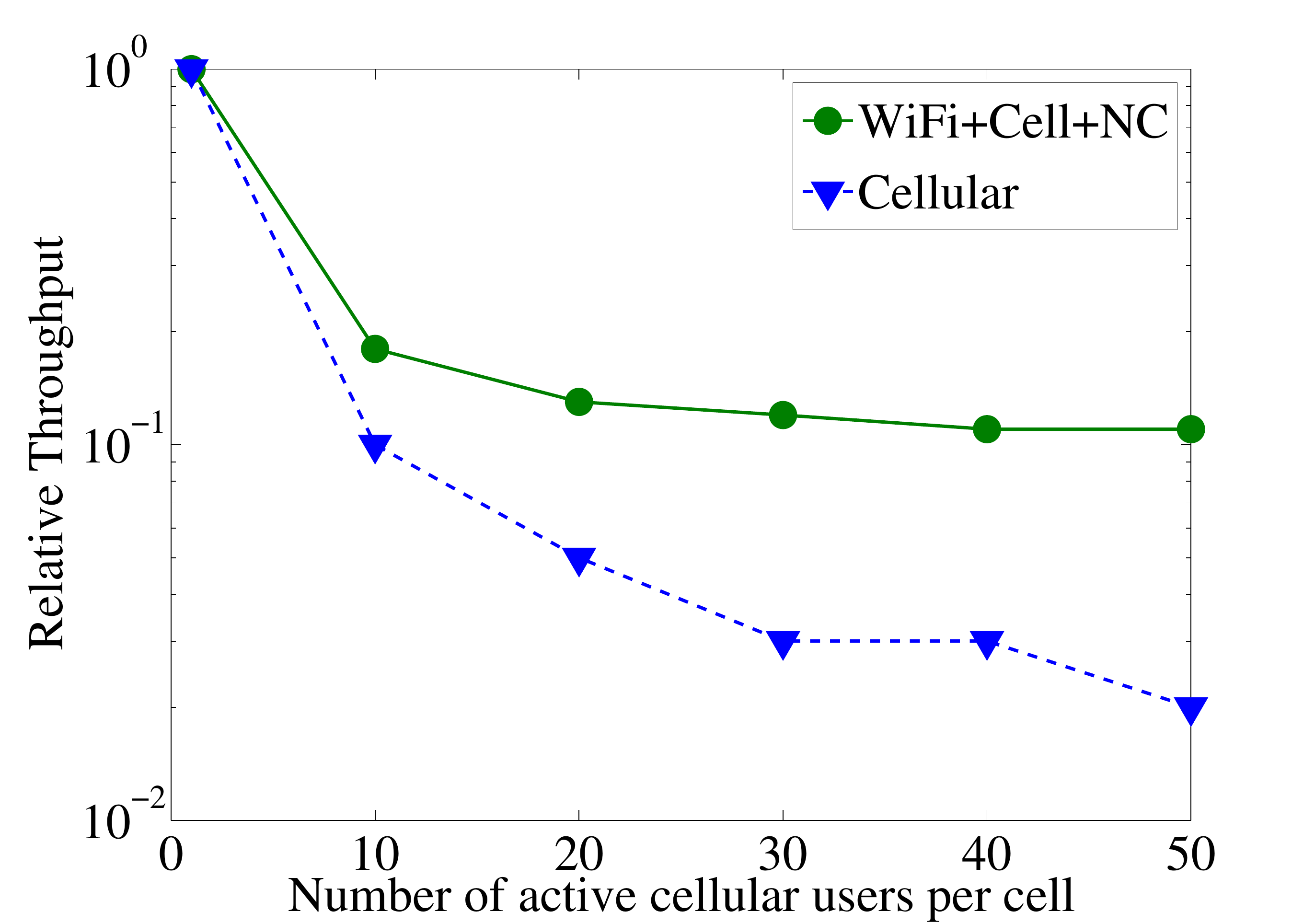}
%\end{center}
\caption{Effect of loading}
\label{fig:load}
\end{figure}

Figure \ref{fig:load} shows the effect of loading on the cellular network due to the presence of multiple cellular users, for a case where the available cellular data rate is equally divided among all users. As the number of cellular users per cell increases in the system, the effective cellular throughput for each user goes down due to loading. But using  network coding with WiFi provides throughput gains that are limited by ad-hoc networks capacity, which is higher than the cellular throughput for large number of users.

\subsubsection{Case 2: Infrastructure based WiFi network}
Thus far we have assumed that the wireless nodes were able to form an ad-hoc network using their WiFi links. However, if we assume the presence of WiFi enabled access to wired infrastructure, i.e., some of the wireless nodes serve as access points with connectivity to a high capacity wired backbone network, then the advantages of HetNetwork coding become even greater. In fact this assumption  extends our system model with seven cell structure to resemble a  real world scenario where WiFi access points can be present in the cellular system. Let us assume that there are a total $n$ nodes per cell in  our system model, and among these $k$ nodes have direct access to a wired backbone network. We simulate communication sessions in this network setting and for varying value of $\frac{k}{n}$ calculate the end-to-end throughput for a S-D pair.
\begin{figure}[t]
%\begin{center}
\includegraphics[width=0.45\textwidth]{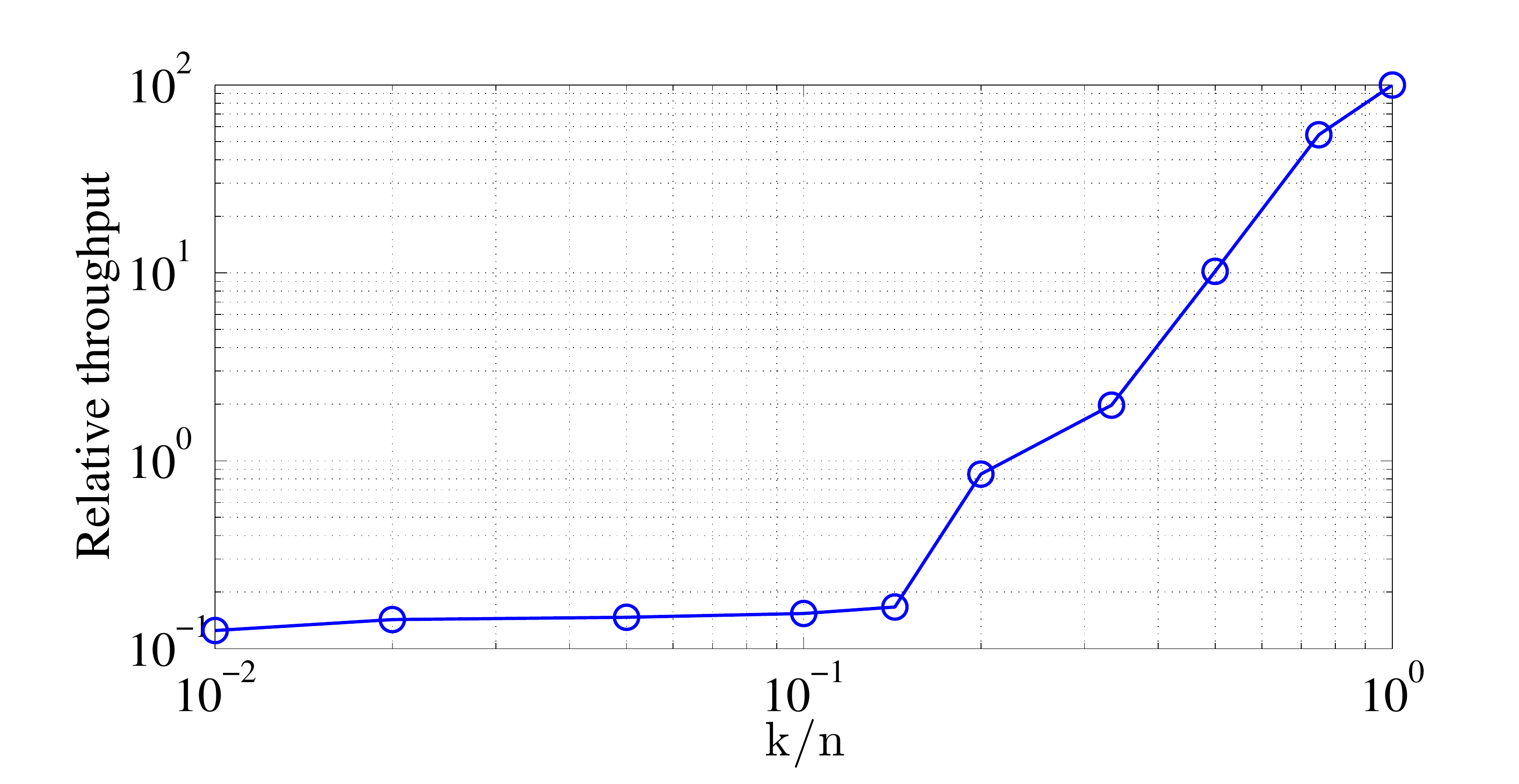}
%\end{center}
\caption{Throughput for infrastructure based WiFi network, when backbone network throughput  $=$ $100\times R_{Wifi}$}
\label{fig:nk}
\end{figure}

Figure \ref{fig:nk} shows the relative throughput for the case when the backbone network throughput is 100 times that of the WiFi link throughput. To interpret this graph, we can see that when $\frac{k}{n} = 1$ (i.e. all nodes are connected to backbone network), all of the nodes can access all other nodes at rate, $100\times R_{WiFi}$. As the fraction of nodes with direct access to the backbone network is decreased, the relative throughput saturates  around the same value as given by the ad-hoc WiFi network in figure \ref{fig:rate}. This is expected because as the number of infrastructure based nodes decrease, a packet has to go through multiple hops. For example we can see that for $k=1$, the average hop distance between this node and any other node in the cell is $> \frac{R}{r}$, which in our case corresponds to $\approx 10$ hops.  However, we note that as the density of nodes in a cell is increased, the curve in figure \ref{fig:nk} shifts to the left, since there is a greater probability of finding a node with backbone network access in it's neighborhood.

\subsection{Experiments on the ORBIT testbed}
To approximate the cellular network from our system model on the ORBIT testbed, we used a WiMAX base-station as a representative  of a cellular base-station. The “profile A” WiMAX (802.16e) base-station used for experimentation is from NEC Corp. and has an outdoor setup that is fully operational with a roof mounted antenna and an FCC experimental license. We used 8 radio nodes in the ORBIT testbed capable of communicating using both WiFi and WiMAX. Specifically, each node contains an Atheros 5000X mini-PCI card for WiFi transmission and an Intel 6250 mini-PCIe 802.11 and 802.16 card for WiFi/WiMAX transmission. There is a configurable RF attenuation matrix between these nodes, which allows us to create various network topologies.

\begin{figure}[t]
\begin{center}
\includegraphics[width=0.3\textwidth]{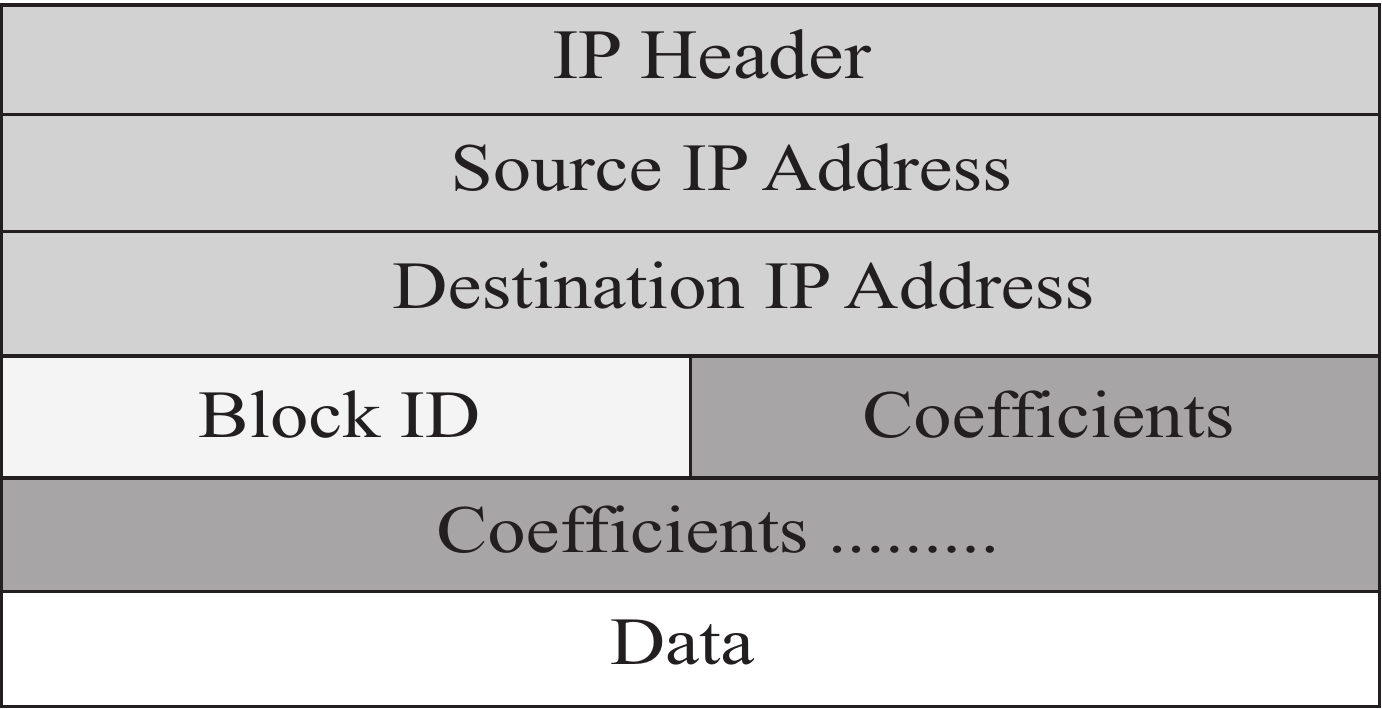}
\end{center}
\caption{IP header used in the ORBIT testbed}
\label{fig:orbit_packet}
\end{figure}

\begin{figure}
        \centering
        \begin{subfigure}[b]{0.43\textwidth}
        \centering
                \includegraphics[width=\textwidth]{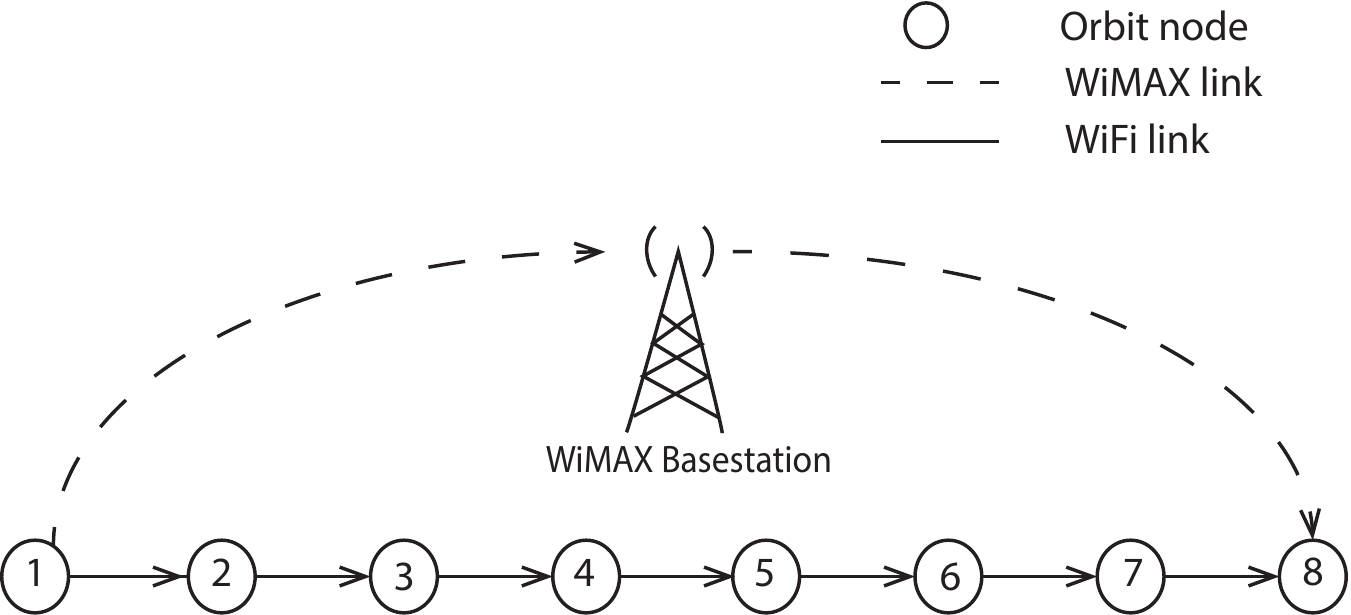}
                \caption{Topology 1}\vspace{1cm}
                \label{fig:topo1}
        \end{subfigure}
        
        \begin{subfigure}[b]{0.45\textwidth}
        \centering
                \includegraphics[width=0.55\textwidth]{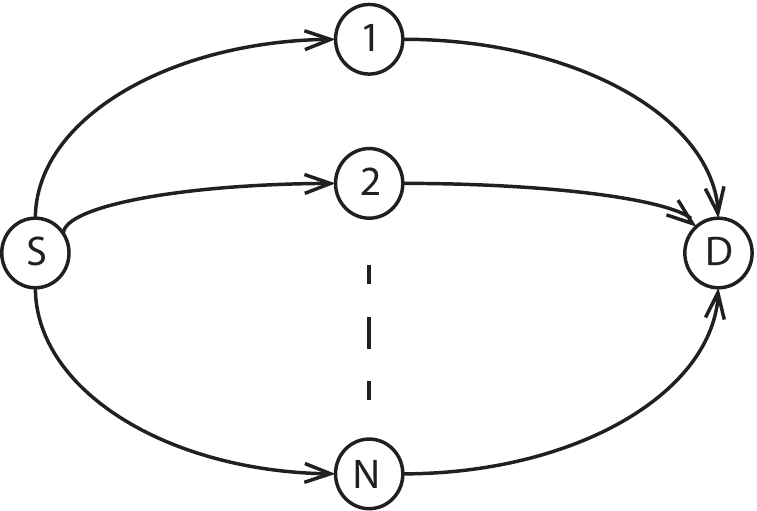}
                \caption{Topology 2}
                \label{fig:topo2}
        \end{subfigure}
        \caption{Topologies used in the ORBIT tesetbed}\label{fig:topos}
\end{figure}

The block size $M$ for the network coding was chosen to be 20.  The network coding coefficients and block ID were added by using the \emph{Options} field of the IPv4 header as shown in figure \ref{fig:orbit_packet}. Each packet's header  remains the same up to the destination IP address section, after which we have 2 bytes for block ID and 30 bytes for coding coefficients.  UDP packets were generated at the source node and routed  to the destination node via a WiMAX base-station and via neighboring nodes with WiFi links. Upon receiving the packet, any node transfers the packet in the direction of the destination node (because of our assumption of an existing routing protocol) after randomly combining all coefficients from the packets in it's memory for this source destination pair. Once the destination node receives sufficient number of packets (i.e. 20 independent set of coefficients) it sends an acknowledgement using the WiMAX channel. Upon receiving the acknowledgement the source starts sending packets from the new block with an updated block ID. We use 2 different network topologies for our experiments as described next.

\subsubsection{Topology 1}
\begin{figure}[t]
%\begin{center}
\includegraphics[width=0.45\textwidth]{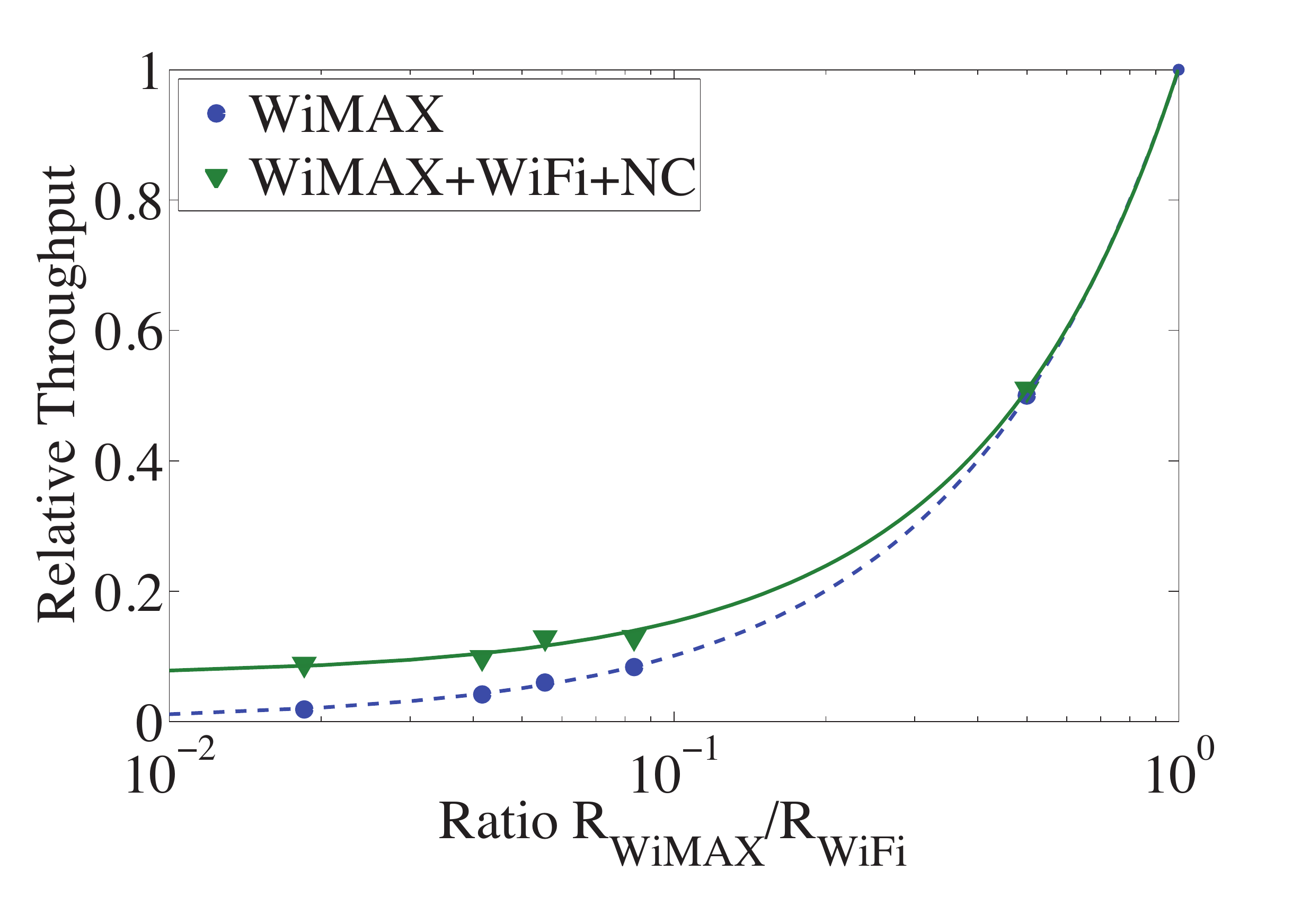}
%\end{center}
\caption{Relative throughput for topology 1}
\label{fig:orbitres1}
\end{figure}

In Topology 1, as shown  in figure \ref{fig:topo1}, we create  two parallel links between node 1 and node 8, one using a direct WiMAX connection, and the other using a  7 hop ad-hoc WiFi link. The WiMAX base station does not perform any coding on the packets. However, the WiFi relay nodes can store some number of packets from the same session and the same block for intra-session network coding.  In this topology, the  WiMAX link has a constant throughput of 1Mbps.   We vary the ratio $R_{WiMAX}/R_{WiFi}$ by varying data rate in the WiFi links and calculate the relative throughput. The relative throughput is calculated  for (a) only WiMAX and (b) combination of WiMAX and WiFi,  normalized by WiFi link throughput. We normalize the throughput to compare the results obtained from the ORBIT testbed to the MATLAB simulation result. Figure \ref{fig:orbitres1} shows the results obtained from the ORBIT testbed, where we observe a  similar trend as the simulation results  given in figure \ref{fig:rate}, but the performance is not as good as the simulations for this topology. This degradation in the  performance is caused by  processing and caching delay at the intermediate nodes, which were assumed to be zero in our simulation. We note that the effect of this delay can be minimized with faster processing and caching, and use of longer block lengths.  However using larger block length will affect the performance of the system when there are requirements on real-time communications, since the destination node has to wait  for more time before it can decode a block.

\subsubsection{Topology 2}
\begin{figure}[t]
%\begin{center}
\includegraphics[width=0.45\textwidth]{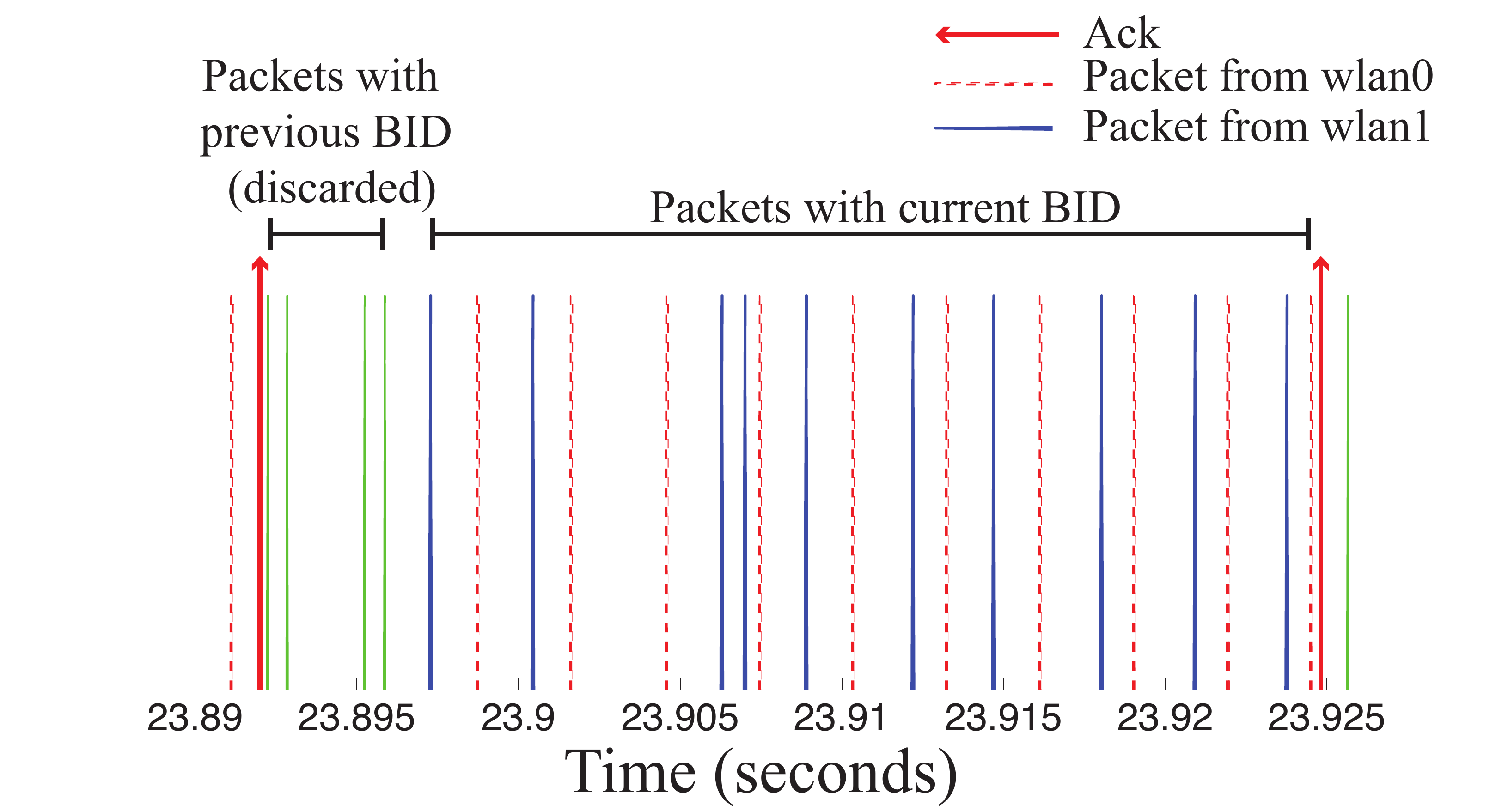}
%\end{center}
\caption{Arrival of Network coded packets at the destination for topology 2}
\label{fig:time}
\end{figure}

\begin{figure}[t]
%\begin{center}
\includegraphics[width=0.45\textwidth]{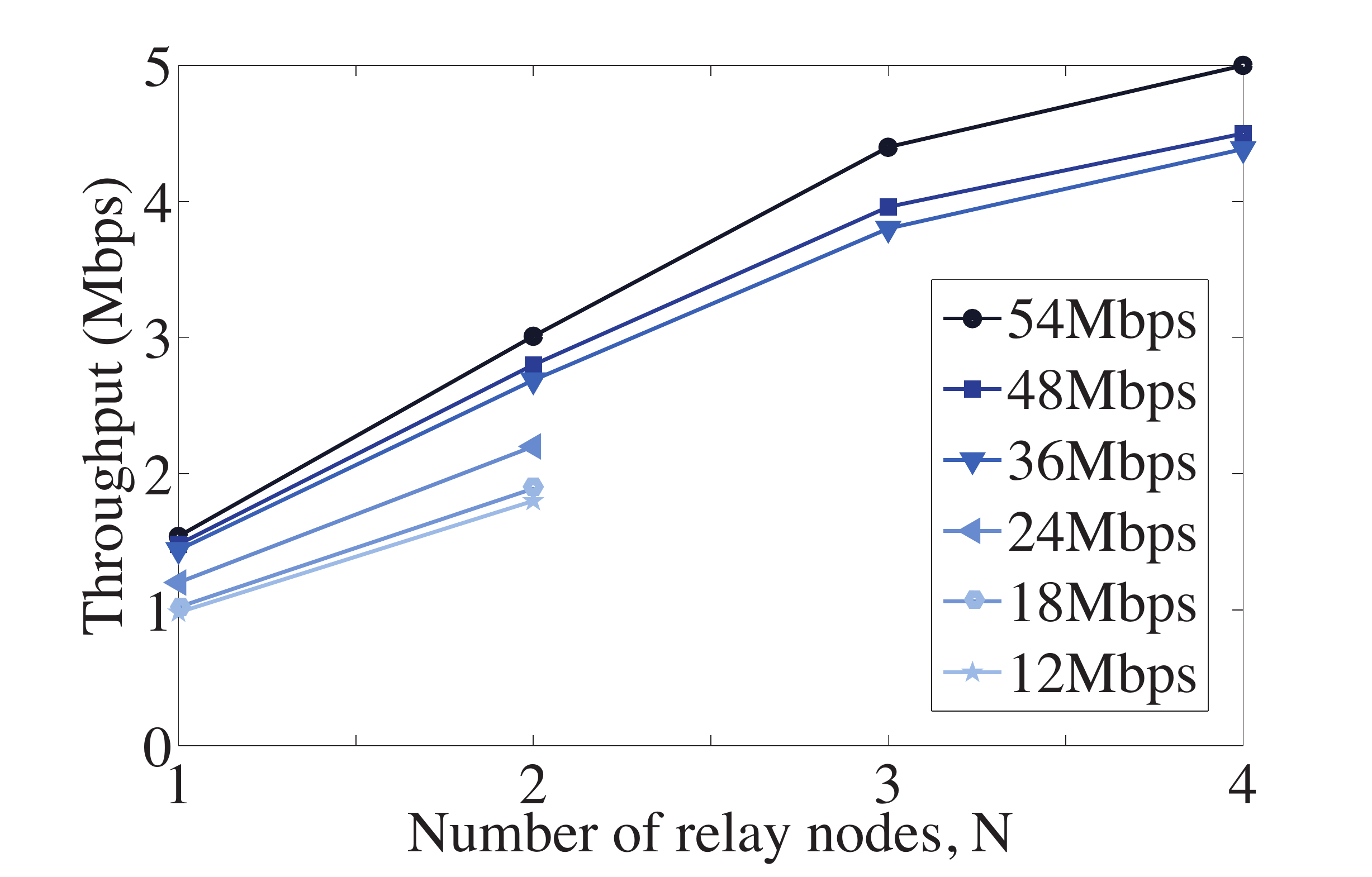}
%\end{center}
\caption{Throughput for topology 2}
\label{fig:orbitres2}
\end{figure}

Figure \ref{fig:topo2} shows the topology where we create a scenario with nodes with multiple WiFi interfaces, where each interface can connect to an access point on a different channel and these access points are each connected to a high capacity link, e.g. an optical fiber link. This topology is also applicable to cases where nodes are allowed to access the network using multiple cellular links simultaneously. Thus this topology represents the situation where it is not possible to form an ad-hoc network between the nodes, but the interfaces can connect to the backbone network via an access point. In the ORBIT testbed, each node has two WiFi interfaces which are used in this experiment. Figure \ref{fig:time} shows the arrival of packets at the destination node for one block of packets (between time 23.89s to 23.925s from the start of experiment), when the data rate for both links are 54Mbps and the packet size is 1200 bytes. This figure shows that once an acknowledgment for a block ID has been received by the source node, it takes some time for the intermediate nodes to gain knowledge about this, and during that period they try to forward packets with  block ID, which has already been acknowledged. Once the intermediate nodes start sending  packets from the current block, then the destination node receives overall $M=20$ packets with current block ID from either of the interfaces. In the particular capture shown in figure 11, the arrivals were linearly independent and therefore the destination node sends an acknowledgement for this block ID. Figure \ref{fig:orbitres2} shows the throughput achieved between  source and  destination for different link capacities and for different number of relay nodes (represented by `N' in figure \ref{fig:topo2}). This figure shows that as the value of N is increased, throughput  increases approximately linearly with N. In this figure, the data set for lower data rates (24, 18 and 12 Mbps) is smaller due to the limitation of only two WiFi links per node in our testbed. While for higher data rates it was possible to share one interface with multiple relay nodes without saturating the channel, that was not the case for lower data rates, but we believe that with the availability of more interfaces, linear scaling of throughput can be observed even for lower data rates. 

%\textcolor{red}{Results for these are not complete/conclusive}
\section{Conclusion}
In this paper we proposed the use of HetNetwork coding  in a heterogenous  network, where nodes have multiple radio interfaces and which can be used in parallel along with network coding to increase the overall throughput. We specifically analyzed this method for nodes with two interfaces (WiFi and cellular). Our simulation results show that when cellular channel is bad compared to WiFi links, our proposed method provides significant gain, even when allowing WiFi links to form an ad-hoc network. We also find that the use of infrastructure based WiFi performs even better. We verified our simulation results by implementing implementing HetNetwork coding on the the ORBIT Testbed with a WiMAX base station and WiFi links. Our experiments show similar results but with caching and processing delay. We find empirical results showing that if the opportunities for multiple parallel paths are available then network coding allows approximately linear throughput gains with number of interfaces.
%\textcolor{red}{To be finished}

\bibliographystyle{IEEEtran}
\bibliography{projref}

\end{document}